\documentclass{mem}
\usepackage{natbib}
\usepackage{txfonts}
\usepackage{balance}
\usepackage{graphicx}
\usepackage[a4paper]{hyperref}
\idline{82}{1}
\begin{document}

\title{
Turbulent Flow and Stirring Mechanisms in the Cosmological Large-scale Structure
}

   \subtitle{}

\author{
L. \,Iapichino}

\institute{
Zentrum f\"ur Astronomie der Universit\"at Heidelberg, 
Institut f\"ur Theoretische Astrophysik, Albert-Ueberle-Str.~2, D-69120
Heidelberg, Germany \\
\email{luigi@ita.uni-heidelberg.de}
}

\authorrunning{Iapichino}

\titlerunning{Turbulence and stirring in the LSS}

\abstract{
Halo mergers and shock waves play a crucial role in the process of hierarchical clustering. Hydrodynamical simulations are the principal investigation tool in this field for theoreticians, and predict that a by-product of cluster formation and virialisation is the injection of turbulence in the cosmic flow. Here I will summarise results from a series of recent works focused on the main stirring mechanisms acting on baryons: minor and major cluster mergers, and curved shocks. Unresolved turbulence has been treated with the implementation of a subgrid scale model.  Recent simulations show that the production of turbulence differs significantly for the warm-hot intergalactic medium (WHIM) and the intra-cluster medium (ICM), because of different stirring mechanisms acting in the two gas phases.

}
\maketitle{}

\section{Introduction}
\label{intro}

The driving of turbulence in the intergalactic medium is a natural consequence of the hierarchical growing of the large-scale structure. Interestingly, turbulence is both linked to the thermal properties of the cosmic baryons (as a by-product of the virialisation mechanism) and to the non-thermal diffuse cluster emission, because of the role which it is expected to play in the acceleration of cosmic rays \citep{fgs08,bl11} and in the amplification of magnetic fields \citep{ssh06,rkc08}.  

The study of turbulence in the framework of the physics of galaxy clusters turns out to be challenging. From an observational viewpoint, important progress has been made by measuring resonant scattering suppression \citep{cfj04,wzc09} and with {\it XMM-Newton} observations of clusters with a compact core \citep{sfs11}. Numerical simulations, on the other hand, have to cope with the big  scale separation between the integral length scale for turbulence injection and the Kolmogorov scale, where the kinetic energy is dissipated by viscous effects. In grid-based hydrodynamical codes the use of adaptive mesh refinement (AMR) can only partially cure this issue; many efforts have been put in designing refinement criteria suitable for turbulent flows \citep{sfh09,in08,vbk09}. Although recent simulations finally reach a fairly large dynamical range \citep[e.g.,][]{vbg11}, it is not possible, even with AMR, to resolve the whole turbulent cascade down to the dissipative length scale. For this reason, turbulence subgrid scale (SGS) models \citep{sb08,mis09} provide  the most physically motivated way of studying the effect of unresolved turbulence on the system (see also \citealt{sf10} for a thorough discussion on this point).

In the following I will briefly summarise some recent results on the study of the main driving mechanisms of turbulence in the cosmic flow at cluster scales: mergers and curved shocks. The work is based on hydrodynamical simulations performed with the ENZO code \citep{obb05}. In this code framework, a subgrid scale model was implemented and coupled both with the equations of fluid dynamics at resolved length scales, and with the AMR \citep{mis09}. The resulting numerical scheme has been called FEARLESS (Fluid mEchanics with Adaptively Refined Large Eddy SimulationS), and combines the adaptive refinement of regions where turbulent flows develop with a consistent modelling of the unresolved turbulence. This tool has proved to be extremely useful in the modelling of turbulent clumped flows, like those in galaxy clusters. The reader is referred elsewhere \citep{mis09,isn11} for numerical details and tests of the SGS model and of FEARLESS. Recent developments and improvements in the modelling of astrophysical turbulence are described by \citet{sf10}.

\section{Simulations of cluster mergers}
\label{mergers}

Galaxy clusters evolve mainly by accretion of smaller clumps; in this process, turbulent kinetic motions are driven in the ICM. In case of minor mergers, the shearing instability develops at the interface between the ICM and the subcluster gas, resulting in the injection of turbulence in the region past the subcluster motion. The problem has been studied by means of high-resolution, idealised hydrodynamical simulations (e.g., \citealt{hcf03,t05a,ias08}), and is relevant for the physics of merger cold fronts \citep{mv07}. 

\begin{figure*}[t!]
\resizebox{\hsize}{!}{\includegraphics[clip=true]{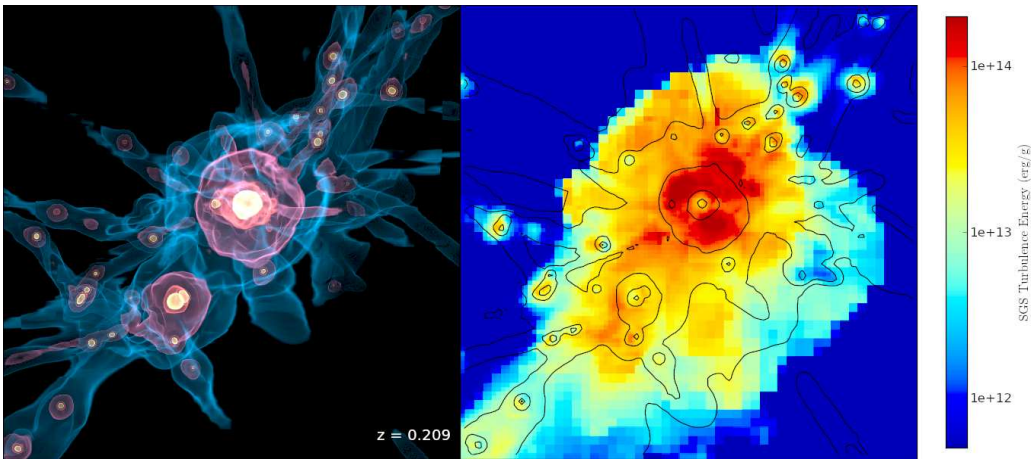}}
\caption{\footnotesize
Visualisation of a region of $12.8\ \mathrm{Mpc}\ h^{-1}$, centred on a cluster ($M = 5.95 \times 10^{14}\ M_{\odot}\ h^{-1}$ at $z = 0$), at $z = 0.209$. In the left-hand panel a volume rendering of the baryon density is reported, whereas in the right-hand panel a projection of the SGS turbulence energy is shown. Simulation details in \citet{mis09}. This is a snapshot from a full movie of the simulation, available at the website {\tt http://www.\-ita.\-uni\--hei\-del\-berg.de/\~{}lui\-gi/\-movies.html}.
}
\label{two-panels}
\end{figure*}

Minor mergers have been also explored in full cosmological simulations of cluster evolution. In \citet{mis09} we used the FEARLESS approach to study the role of merger-induced turbulence in the ICM. Since the flow in the ICM is subsonic, we found that gas kinetic energy (either resolved or SGS) is just a minor part of the cluster energy budget. The production of turbulence is closely correlated with mergers: this is clearly visible in Fig.~\ref{two-panels}, where a projection of the SGS turbulence energy (computed using FEARLESS) is compared with a volume rendering of the baryon density, showing the cluster substructure. For example, one can observe the small subcluster immediately on the left of the cluster core, which is moving downwards around the centre and has stirred the ICM in its turbulent wake, as indicated by the large value of the SGS turbulence energy in that region. The dissipation of unresolved turbulence results also in a larger core temperature and a higher core entropy.

\begin{figure}[]
\resizebox{\hsize}{!}{\includegraphics[clip=true]{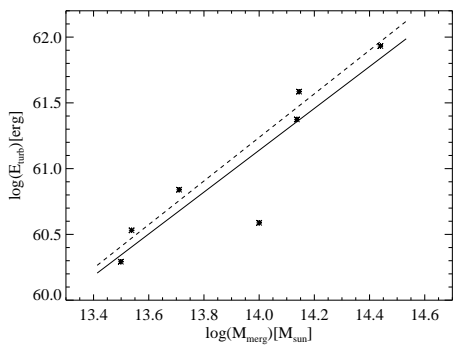}}
\caption{
\footnotesize
The turbulence energy is plotted against the merger mass $M_{\mathrm{merg}}$, for the merging clusters of the sample of \citet{pim11}. The solid line is the best fit to the data points, corresponding to a scaling law $E_{\mathrm{turb}} \propto M^{1.6}$, whereas the dashed line (with a slope of 1.66) is the best fit computed by excluding an outlier cluster (the point at $\log M = 14.0$).
}
\label{scaling}
\end{figure}

The importance of mergers and merger-induced turbulence for the cluster energy budget is apparent from the study of major merger simulations performed by \citet{pim11}. From the study of a sample of mergers, it was found that the ratio of the turbulent to total pressure in the cluster core is larger than $10\%$ for about $2\ \mathrm{Gyr}$ after a major merger. The scaling of the turbulence energy with the cluster mass (Fig.~\ref{scaling}) in the sample is consistent with $M^{5/3}$, which is the same scaling expected for the thermal energy in the self-similar model. This result highlights again that virialisation and turbulence injection are two faces of a same physical process, the hierarchical formation of cosmological structure.

\section{Turbulence production in the intergalactic medium}
\label{igm}

There are several mechanisms that are able to stir the baryons and inject turbulence in the cosmic flow at cluster scales. In the previous Section, cluster mergers were presented. If the effects of galaxy motions in the IGM and AGN outflows are neglected, a remaining stirring mechanism is provided by the baroclinic vorticity generation \citep{krc07}. In filaments and cluster outskirts the unprocessed gas is accreted and shock-heated in the forming structures, and the injection of turbulence is a by-product of this gas accretion at curved shocks.

Both stirring mechanisms were studied in detail in \citet{isn11}, by means of a FEARLESS simulation in a cosmological box with a volume of $(100\ \mathrm{Mpc}\ h^{-1})^3$. The analysis was focused on baryons with temperature larger than $10^5\ \mathrm{K}$, and a distinction was done between gas in collapsing structures with baryon overdensity $\delta > 10^3$ and less dense material. For consistency with previous studies in this field, it was chosen to refer to these two baryon phases as to ICM and warm-hot intergalactic medium (WHIM), respectively.

\begin{figure}[]
\resizebox{\hsize}{!}{\includegraphics[clip=true]{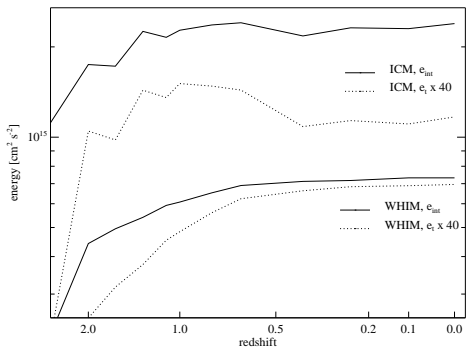}}
\caption{
\footnotesize
Time evolution of the mass-weighted averages of specific internal (solid lines) and SGS turbulent (dotted lines) energies, for the two baryon phases under investigation. The two lines in the upper part of the plot refer to the ICM, and the other two to the WHIM. The lines are scaled according to the factors in the legends, in order to be accommodated in the same plot.
}
\label{whim-icm}
\end{figure}

In Fig.~\ref{whim-icm} the time evolution of the specific internal and SGS turbulent energies is shown for both baryon phases. The production of turbulence has clearly a different redshift dependence in the ICM and WHIM. We argue that this difference is due to the mechanisms of turbulence generation that are dominant in the two baryon phases.

In the ICM, turbulence is produced mainly by merger events: for this gas phase, the broad peak of $e_\mathrm{t}$ at redshift between 1.0 and 0.65 is consistent with the major merger phase of halos in the mass range $10^{13}\ M_\odot < M < 10^{14}\ M_\odot$ \citep[for instance,][]{gms07}. For the WHIM gas, residing in cluster outskirts, smaller clumps and filaments, the evolution of turbulence is related to the amount of kinetic energy processed by the external shocks \citep{soh08}, where turbulence is injected by the baroclinic mechanism. For further details, and for a discussion about the effects of dynamical pressure support on the gravitational contraction of the gas, we refer the reader to \citet{isn11}.

\begin{acknowledgements}
The ENZO code is developed by the Laboratory for Computational Astrophysics at the University of California in San Diego ({\tt http://lca.ucsd.edu}). The data analysis was performed using the {\tt yt} toolkit \citep{tso11}. Thanks to my collaborators involved in the projects mentioned above, in particular to J.~Niemeyer, W.~Schmidt and S.~Paul.
\end{acknowledgements}

\bibliography{cluster-index}
\bibliographystyle{bibtex/aa}

\end{document}